# "Shadowing" of the electromagnetic field of relativistic charged particles

G Naumenko[1], X Artru[2], A Potylitsyn[3], Yu Popov[3], L Sukhikh[3] and M Shevelev[1]

[1]Nuclear Physics Institute of Tomsk Polytechnic University, Lenina str. 2a Tomsk, 634050, Russia

[2]Institut de Physique Nucl.eaire de Lyon, Université de Lyon,CNRS- IN2P3 and Université Lyon 1, 69622 Villeurbanne, France

[3]Tomsk Polytechnic University, Lenina str. 2, Tomsk, 634050, Russia

x.artru@ipnl.in2p3.fr

**Abstract**. In radiation processes such as a transition radiation, diffraction radiation, etc. based on relativistic electrons passing through or near an opaque screen, the electron self-field is partly shadowed after the screen over a distance of the order of the formation length $\gamma^2\lambda$. This effect has been investigated on coherent diffraction radiation (DR) by electron bunches. Absorbing and conductive half-plane screens were placed at various distances $L$ before a standard DR source (inclined half-plane mirror). The radiation intensity was reduced when the screen was at small $L$ and on the same side as the mirror. No reduction was observed when the screen was on the opposite side. It is worth noting that absorbing and conductive half-plane screens produce the same shadowing effect. The shadowing effect is responsible for a bound on the intensity of Smith-Purcell radiation.

## 1. Introduction

Macroscopic Maxwell equations may be used (in principle) for calculating the radiation emitted by relativistic electrons passing through or near material targets. Nevertheless phenomenological concepts like *equivalent photons* and *formation zone* are useful for an intuitive understanding of the main features. In transition radiation (TR), the *formation zone* is a region of length $l_f \sim \gamma^2\lambda$ where the Coulomb field of the fast electron and the forward TR from the target cannot be measured separately (in this paper we assume that the electron is ultra-relativistic : $\gamma \gg 1$). The total electromagnetic field in this region is reduced by a destructive interference between the two fields. This effect can also be interpreted with the *equivalent photons* method [1] (the terms *virtual quanta*, *quasi-real photons* or *pseudo-photons* are also used). In this picture the target intercepts the beam of equivalent photons, leaving a *shadow* downstream. It takes a length $\sim l_f$ for the electron to re-form its pseudo-photon cloud. This explains in a simple way the suppression of Optical TR in a Wartski interferometer when the distance between the two foils is small compared to $l_f$.

The similar concept of *half-naked* electron was introduced in [2,3] in the physics of electron scattering on atoms or nuclei. These authors consider the Landau-Pomeranchuk effect [4] and the density effect of Ter-Mikaelian [5] as experimental evidences for the half-naked electron effect.

In this paper it will be shown, both experimentally and theoretically, that a similar shadow region exists behind a Diffraction Radiation (DR) target (either conductive or absorbing). However, unlike for TR, the shadow is only on one side of the particle trajectory (the same side as the target), as expected from a naive application of optical geometry to the pseudo-photons. In this region radiation by a second DR target is inhibited.

**2. Shadowing effect from different points of view**

Here we present different points of view about the shadow effect. Let us consider a fast electron moving near two targets like in figure 1.

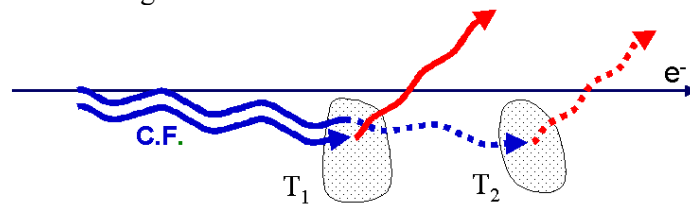

Figure 1. Shadowing of the target by another target.

In the first point of view the Coulomb field (C.F.) is considered as a beam of quasi-real photon. Scattering of this field by the first target gives Diffraction Radiation. The second target is in the shadow of the first one, therefore emits almost no radiation. The Coulomb field is gradually "repaired" during the formation zone of length $l_f \sim \gamma^2 \lambda$.

A precise space-time description of a "half-naked electron", in the context of the Landau-Pomeranchuk effect mentioned above, is given in [3], where figures 1.1 and 2.4 depict the restoration of the Coulomb field. In figure 2 we adapted this picture to the case of a particle passing through a narrow hole.

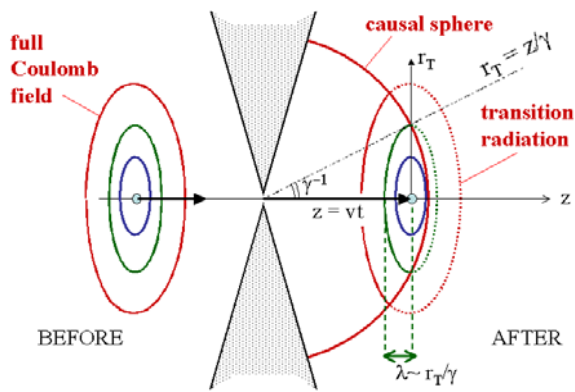
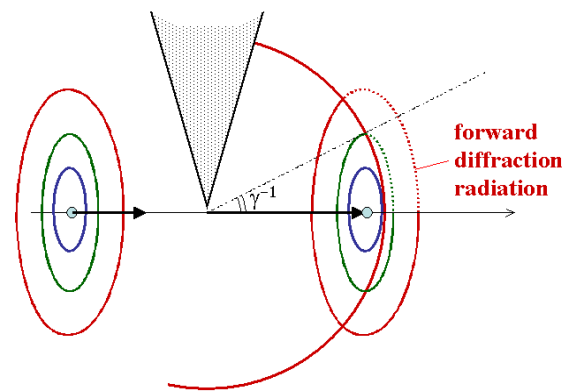

Figure 2. Field of a particle passing through a hole (analogous to figure 1.1 of [3]).

Figure 3. Shadowing by a half-screen.

On the right-hand side, according to the relativistic causality principle, no field exists outside the expanding sphere $|\mathbf{r}| \leq t$. Inside, the field is fully restored. Outside the cone $|\mathbf{r}_T| = z/(\gamma v) \simeq z/\gamma$, there is no field before the electron arrives at the same $z$, therefore the Coulomb field is more than half-amputed. Taking the temporal Fourier decomposition of the Coulomb field, we have $\omega \leq \gamma v / |\mathbf{r}_T|$, therefore the condition for a good restoration of the frequency $\omega$ is $z \geq \gamma |\mathbf{r}_T|$ for all $|\mathbf{r}_T| \in [0, \gamma/\omega]$. Up to a factor $2\pi$, we recover the condition $z \geq l_f$.

Figure 3 depicts an analogous situation where the particle passes near the border of a screen. In this case only one side of the Coulomb field is removed. The goal of the experiment presented below is to observe this one-sided shadow region.

The second point of view has been illustrated by B.M. Bolotovskii in figure 4 [6]. The traveling Coulomb field (represented by ellipses centered on the successive positions of the particle) induces current in the half-plane screen, which in turn emit Diffraction Radiation (DR), represented by small pieces of ellipses. The radiation field is such that close to the screen it kills part of the particle field. In this point of view, shadowing appears as a destructive interference effect between the Coulomb field, $CF$, and the forward diffraction radiation field, $FDR$.

The interference gradually disappears (positions 3, 4, 5 of the figure) due to the different velocities, $v \simeq 1 - \gamma^{-2}/2$ and $c = 1$ (in our units), of the Coulomb and radiation fields. These fields get out of phase after a time $t_f \sim \lambda/(c-v) \sim l_f$.

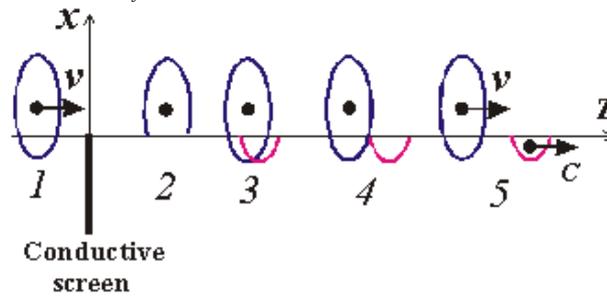

Figure 4. Illustration of the radiation formation length effect by
B.M. BolotovskiI in [6].

In this second point of view, the target $T_2$ of Fig.1 is submitted to the usual Coulomb field of the particle plus the field of the forward diffraction radiation from $T_1$, denoted $FDR^{(1)}$. Each of these fields induces a current in $T_2$ and this current generates a new radiation. Outgoing from $T_2$, we have then the superposition of
- the ordinary diffraction radiation from $T_2$, denoted $DR^{(2)}$,
- a part of FDR from $T_1$ rescattered by $T_2$, denoted $R_2 \{FDR^{(1)}\}$.

If $T_2$ is at distance less than $l_f$ from $T_1$, the total inducing field, the total induced current and the sum $DR^{(2)} + R_2 \{FDR^{(1)}\}$ are suppressed. In this viewpoint, illustrated in Fig. 5, the inhibition of radiation from the second target is a combined result of rescattering and destructive interference.

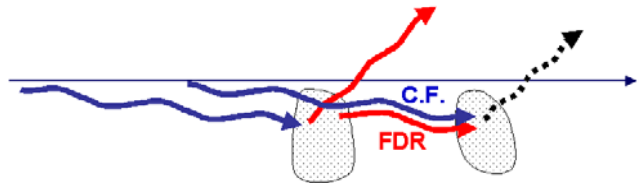

Figure 5. Shadow effect as a re-scattering effect.

## 3. The Tomsk experiment

In the Landau-Pomeranchuk effect the formation zone is not directly observed. In optical transition radiation, it has a macroscopic longitudinal size. However its transverse size $\sim \gamma\lambda$ is usually small compared to the width of the electron beam (this is the effective transverse size of the Coulomb field at observed wavelength $\lambda$). Therefore a direct investigation of its transverse properties needs higher values of the product $\gamma\lambda$.

*3.1. Principle of the experiment*

The Tomsk microtron delivers an electron beam with a Lorenz-factor $\gamma \simeq 12$. The radiation is observed in the microwave domain ($\lambda \sim 1$ cm). Thus the transverse scale $\gamma\lambda$ is about 12 cm, which is quite comfortable. The formation length $\gamma^2\lambda$ is about 1.4 m.

The energy of the radiation coming from the interaction of one electron with the targets, in the bandwidth $\Delta\omega$, is typically of the order of $\alpha\hbar\Delta\omega$. In the microwave region, it is too small to be measurable, even if we multiply it by the number $N_e$ of electrons in a bunch. However, if the bunch length $\sigma_z$ is smaller than $\lambda$ the radiation has a coherent character which makes it proportional to $N_e^2$ rather than $N_e$. More precisely, the radiation intensity from the bunch is given by

$$I_{bunch} = \left[ N_e(N_e - 1) f^2(\lambda/\sigma_z) + N_e \right] I_e$$

where $f(\lambda/\sigma_z)$ is the longitudinal form-factor of the bunch. In this experiment, $N_e \sim 10^8$, therefore the bunch coherence amplifies the radiation by 8 orders of magnitude, making it easily observable.

Shadowing is achieved by the reflection of the pseudo-photons by a half-plane mirror or their absorption in a half-plane screen. We can indeed assume that pseudo-photons are reflected in the mirror or absorbed in absorber almost like real photons. In particular, it does not matter whether we use an absorber or a mirror for shadowing. This is supported by the analysis of diffraction radiation from a "black body" absorber in [7], although this reference does not analyze the electromagnetic field inside the radiation formation zone.

The electron field in the shadowing region is probed with a second diffraction radiation target, from which one measures the intensity and angular distribution of backward diffraction radiation (BDR), according the scheme shown in figure 6. BDR is easier to measure than FDR, since the separation from the Coulomb field occurs very close to the target.

For the case where targets are on opposite sides of the beam we keep the same $T_2$ target but over-turn $T_1$ to the opposite side (see figure 7).

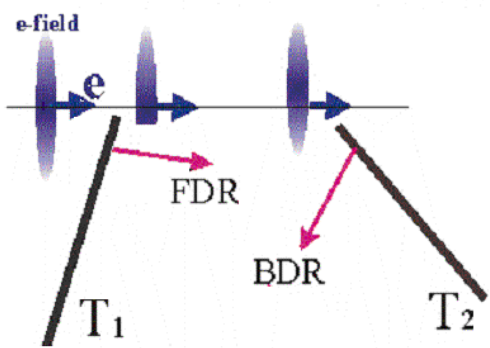 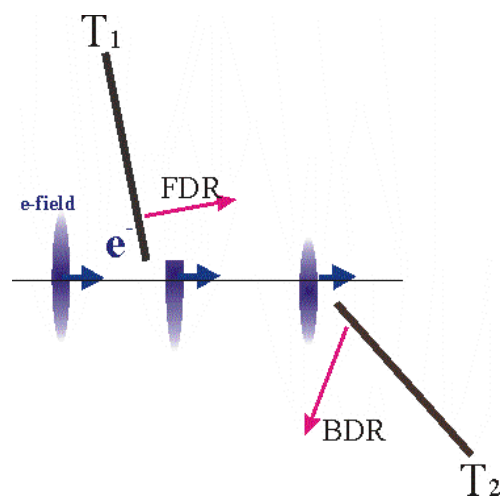

Figure 6. Experimental scheme for the investigation of shadowing. $T_1$ is an absorber or conductive target, $T_2$ is a conductive mirror.

Figure 7. Scheme with the targets on opposite beam sides

*3.2. Experimental setup*

The experiment was performed on the extracted electron beam of the microtron of the Tomsk Nuclear Physics Institute. The beam is extracted from the vacuum chamber through a 20 $\mu$m thick beryllium foil. Its parameters are listed in table 1.

**Table 1.** Electron beam parameters.

| | | | |
|---|---|---|---|
| Electron energy | 6.1 MeV ($\gamma = 12$). | Bunch period | 380 psec |
| Train duration | $\tau \approx 4$ $\mu$sec | Bunch population | $N_e$=6·10$^8$ |
| Bunches in a train | $n_b$≈1.6·10$^4$ | Bunch length | σ≈1.3~1.6mm |

The window caused a beam divergence ($\simeq 0.08$ radian). For this reason the distance $L$ between targets $T_1$ and $T_2$ was limited to $L_{max}$ = 220 mm. At larger distance, it would be necessary to increase the impact parameter $b_2$ of $T_2$, to prevent direct interactions with the side of the beam. This would strongly reduce the diffraction radiation yield. $L$ was changed by moving the upstream target $T_1$, keeping the diffraction radiation mirror $T_2$ fixed. In this way, the influence of a transversal bunch form-factor on the DR yield from was $T_2$ avoided.

The beryllium window with diameter of 34 mm may be also considered as a source of a transition radiation. A simple estimation showed that the intensity of this radiation is negligible downstream $T_1$.

The coherent radiation intensity for $\lambda > 9$ mm is by 8 orders larger than the incoherent one and has a power level =1 Watt per steradian. It means that one can investigate coherent radiation in this wavelength range without difficulty.

For the radiation measurements we used the room temperature detector DP20M, with parameters described in [8]. Its main elements are a low-threshold diode, a broadband antenna and a preamplifier. The detector efficiency in the wavelength region $\lambda$=3~16 mm is estimated to be constant to a ±15% accuracy. The detector sensitivity is 0.3 V/mWatt. A wave-guide with a cutoff $\lambda_{cut}$=17 mm was used to cut the long-wave background of the accelerator RF system. The high frequency limit of the wavelength interval is defined by the bunch form-factor. This limit ($\lambda_{min}$=9 mm) was measured using discrete wave filters [9] and a grating spectrometer.

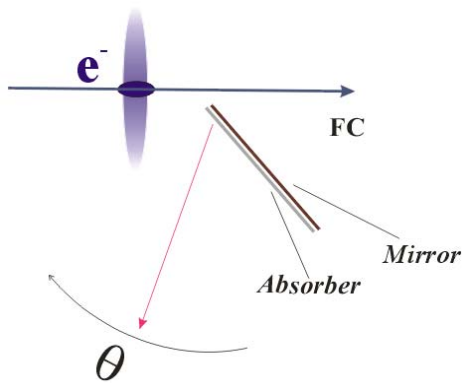
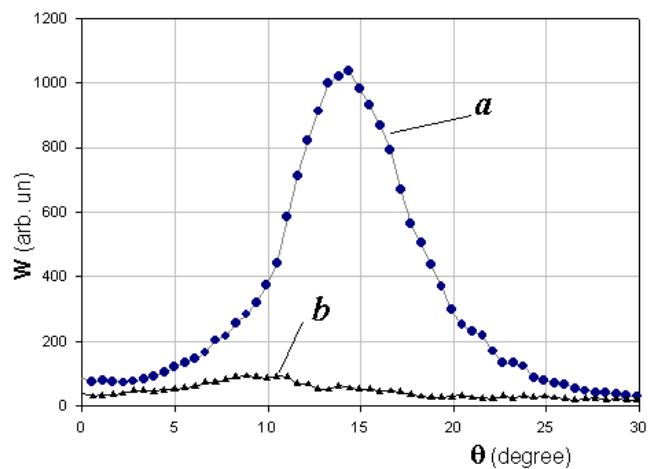

Figure 8. Scheme of the absorber test.

Figure 9. Angular distribution of the radiation intensity, measured in the absorber test (figure 8). *a* – BDR from conductive target without absorber, b – radiation from the same target covered by absorber.

To exclude the prewave zone effect (see [10]) a parabolic telescope was used to investigate the angular distribution. This method was suggested and tested in [11] and gives the same angular distribution as in the far field zone ($R \gg \gamma^2\lambda$).

The absorber properties were checked both with a real photon beam from a source of wavelength $\lambda$ =6 mm and with the measurement of reflected pseudo-photons of beam electrons, according to the scheme shown in figure 8. In figure 9 curve *a* is the BDR angular distribution from a conductive target without absorber and curve *b* is for the same target covered by the absorber. We can see that almost no pseudo-photon reflection is registered. The test with the real photon beam showed that, within $\simeq 3\%$ experimental accuracy, no real photons passed through the absorber and no photons were reflected by it.

### 3.3. Experimental results

Using the described method, the angular distributions of backward radiation from the conductive target $T_2$ were measured for inter-target distances $L$=0 to 220 mm with steps of 20 mm. Two configurations of the experiment were used :

- Configuration *a* (figure 10*a*) is intended to demonstrate the shadowing of the electron field. In order not to restrict the measured angular range of the radiation from $T_2$ at small values of L, the absorber was inclined at 45° as shown in figure 10.
- In configuration *b* (figure 10*b*) the absorber is placed on the side opposite to $T_2$ with respect to the electron beam. In this geometry the shadow effect is expected not to be observed.

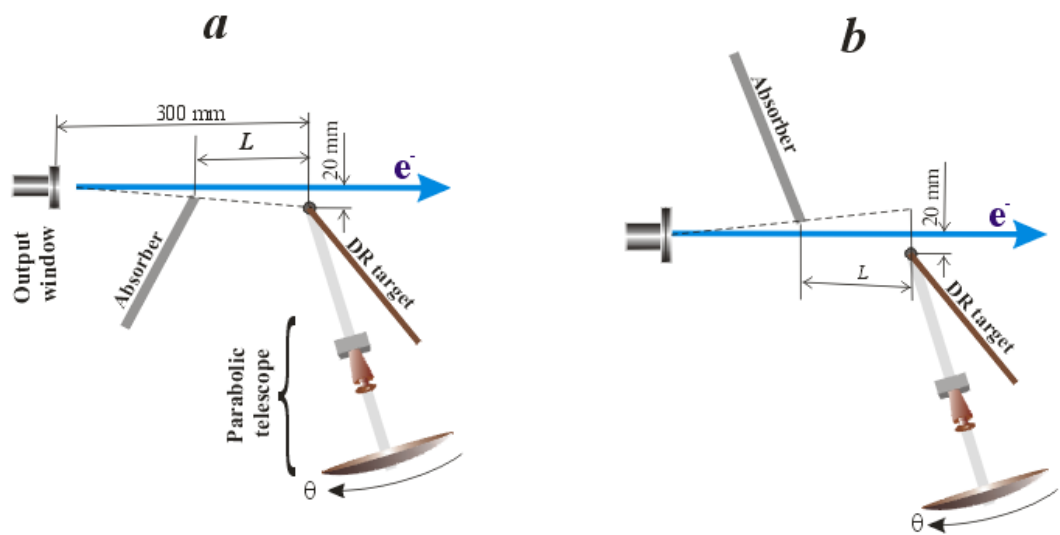

Figure 10. Scheme of experiment on the shadow effect observation. *a* – the targets are on the same side of the electron beam. *b* – the targets are on the opposite sides of the electron beam.

Samples of the measured distributions using configuration ***a*** are shown in figure 11 and the simultaneous dependence on the observation angle $\theta$ and on the distance $L$ is shown in figure 13*a*. The corresponding results for the configuration ***b*** (opposite-side screens) are presented in figures 12 and 13*b*. Figures 11 and 12 have the same units of radiation intensity.

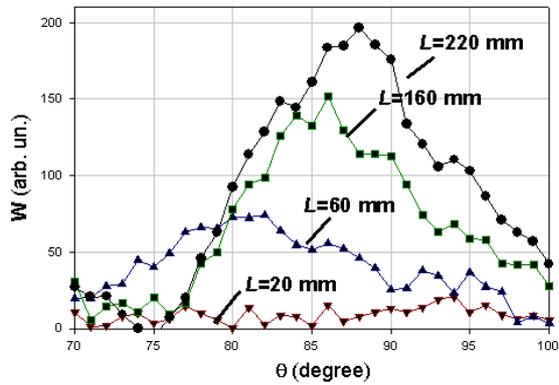
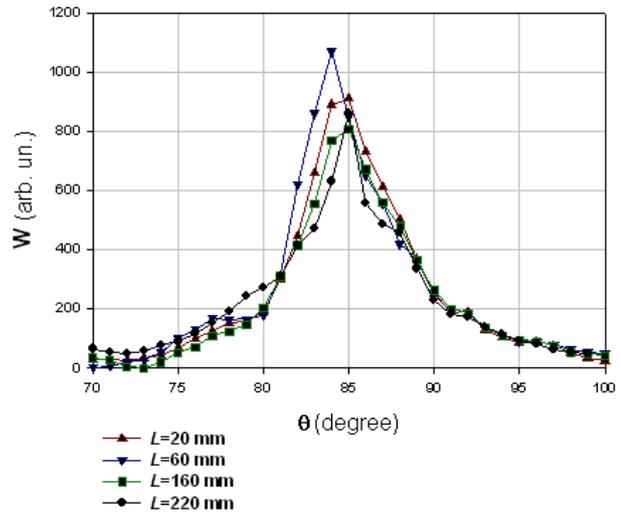

Figure 11. Samples of the measured angular distributions for different distance to the absorber using scheme *a* (figure 10*a*).

Figure 12. Samples of the measured angular distributions for different distance to the absorber using scheme *b* (figure 10*b*).

Figures 13*a* and 13*b* have the same intensity scale to emphasize the shadowing effect. These figures exhibit the qualitative properties of shadowing :
- the intensity in scheme *a* is much mower than in scheme *b*,
- it decreases together with *L*. By contrast, no systematic dependence on *L* is seen in scheme *b*.

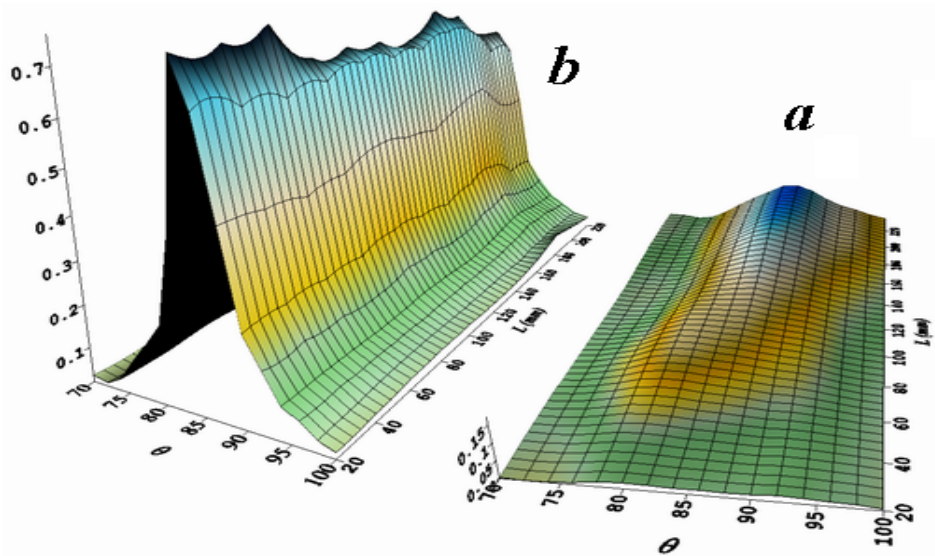

Figure 13. Measured dependence of the radiation intensity on the observation angle θ and the distance L to the absorber. Plots a and b correspond to the configurations a and b of figure 10.

To show that the shadow effect does not depend on the type of the target, the experiment was repeated replacing the absorber target $T_1$ by a conducting screen. The double dependence on $\theta$ and *L* is shown in figure 14. Comparison with figure 13 confirms the independence of the shadow effect on the target type.

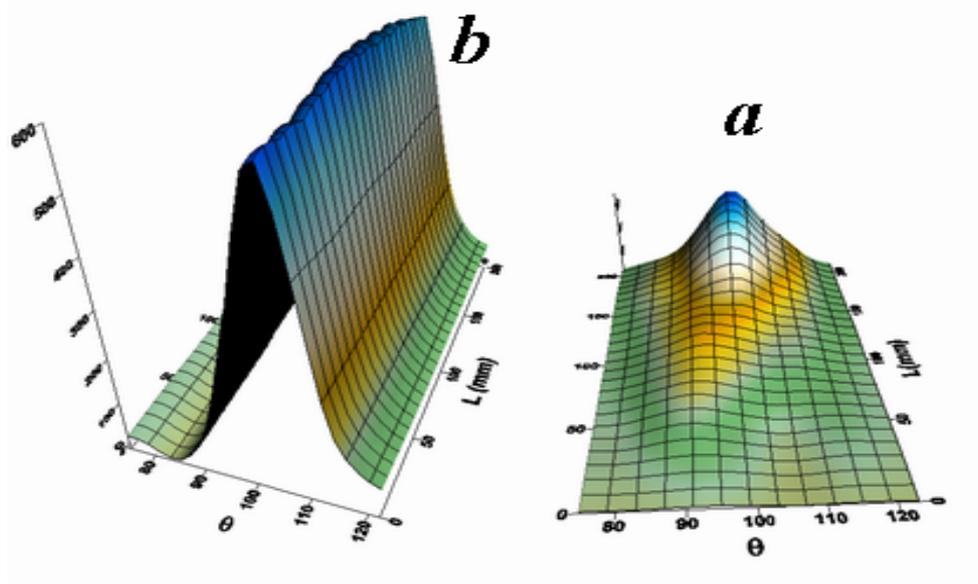

Figure 14. Measured dependence similar to the one shown in figure 13, using a conducting instead of an absorbing target $T_1$.

## 4. Theoretical calculation and comparison with experiment

The calculation of the expected angular distribution of the radiation from the conductive target $T_2$ (see figures 6 and 7), for one electron, can be done with the method outlined at the end of section 2, *i.e.*, express the amplitude as $DR^{(2)} + R_2\{FDR^{(1)}\}$. The results presented below are derived in the ultra-relativistic approximation for the electron ($\gamma \gg 1$) and the paraxial approximation for the photon ($k_T \ll k_z$).

### 4.1. Free field evolution in $z$

In the paraxial approximation of the Huyghens-Fresnel formula, the evolution of a free field from a transverse plane at $z_0$ to a transverse plane at $z$ can be written as

$$\mathbf{E}_{\text{free}}(z) = G(z - z_0)\mathbf{E}_{\text{free}}(z_0), \tag{1}$$

In this symbolic formula, $\mathbf{E}(z)$ represents the field state in the transverse plane of abscissa $z$ at fixed frequency $\omega$. In the $(x, y)$ representation, it stands for $\mathbf{E}(\omega, x, y, z)$ considered as function of the bi-vector $(x, y)$. $\omega$ is considered as a parameter and $z$ plays the role similar to time in quantum mechanics. By a 2-dimensional Fourier transform, one passes to the $\mathbf{k}_T$-representation $\mathbf{E}(\omega, k_x, k_y, z)$. The *propagator* $G$ is an operator. $G\mathbf{E}$ is an ordinary product in the $\mathbf{k}_T$ representation and a convolution product in the $\mathbf{r}_T$ representation. In the $\mathbf{k}_T$ representation,

$$G(\mathbf{k}_T; z) = \exp\left\{i\left(\omega - \frac{k_T^2}{2\omega}\right)z\right\}. \tag{2}$$

In the mixed $(x, k_y)$ representation,

$$G(x, k_y; z) = (iz\lambda)^{-1/2} \exp\left\{i\omega z + i\omega\frac{x^2}{2z} - i\frac{k_y^2}{2\omega}z\right\}. \tag{3}$$

Here $G\mathbf{E}$ is an ordinary product in $k_y$ and a convolution product in $x$.

In the limit $z = +\infty$, Eq.(1) gives

$$\mathbf{E}_{\text{free}}(\omega, \mathbf{r}_T, z_0 + Z) \simeq \frac{\omega}{2\pi Z} \exp\left\{i\omega\left(Z + \frac{r_T^2}{2Z}\right)\right\} \mathbf{E}_{\text{free}}(\omega, \mathbf{k}_T, z_0)|_{\mathbf{k}_T = \omega \mathbf{r}_T/Z} \quad (4)$$

with $Z = z - z_0$. This formula relates the asymptotic angular distribution to the $\mathbf{k}_T$ representation at finite $z_0$.

### 4.2. Coulomb field
The Coulomb field of an electron fast moving along the $z$-axis is[1]

$$\mathbf{E}_C(t, x, y, z) \simeq \frac{-e}{4\pi} \gamma \left[\mathbf{r}_T^2 + \gamma^2(z - vt)^2\right]^{-3/2} \mathbf{r}_T \quad (5)$$

We have neglected the longitudinal component. Using partial Fourier transformation, it can also be represented in energy, transverse momentum and $z$ space,

$$\mathbf{E}_C(\omega, \mathbf{k}_T, z) = \frac{ie}{v} \frac{\mathbf{k}_T}{\mathbf{k}_T^2 + q_0^2} e^{i\omega z/v}, \quad (6)$$

or in energy, $x, k_y$ and $z$ space,

$$\mathbf{E}_C(\omega, x, k_y, z) = \frac{ie}{2v} e^{-\mu|x|} e^{i\omega z/v} (i\,\text{sign}(x), \tau), \quad (7)$$

with $q_0 = \omega/(\gamma v)$, $\mu = \sqrt{k_y^2 + q_0^2}$, $\tau = k_y/\mu$.

The Coulomb field is made of virtual, or *bound* photons, of phase velocity $\omega/k_z = v$, whereas diffraction radiation is made of real photons, *i.e.*, $\omega/|\mathbf{k}| = 1$. If the electron is suddenly stopped, or scattered at large angle, at $z = z_0$ the field (5-7) is "released" and propagates as a free field downstream the $z_0$ point.

### 4.3. Diffraction Radiation from one target
The forward diffraction radiation field $\mathbf{E}_{\text{FDR}}$ from a target T can be considered as the negative of the intercepted Coulomb field, propagating freely in the forward half-space:

$$\mathbf{E}_{\text{FDR}}(z) = -G(z - z_t) A \mathbf{E}_C(z_t), \quad (8)$$

where $z_t$ is the point where the $z$ axis crosses the target plane and $A(x, y)$ is the supporting function of the transverse area of the target. We put the screen in the $x \geq b$ half-space, where $b$ is the impact parameter, therefore $A(x, y) = A(x) = \theta(x - b)$ ($A\mathbf{E}$ is an ordinary product in $(x, y)$ space). Forward transition radiation is obtained in the limit $b = -\infty$. $\mathbf{E}_{\text{FDR}}$ is independent on the target material (once it is fully opaque), on the surface roughness and is unchanged when the screen is tilted. The field after the screen is the sum of FDR and the full Coulomb field $\mathbf{E}_C$ of (5-7).

For a mirror target, the backward diffraction radiation field $\mathbf{E}_{\text{BDR}}$ is symmetrical of $\mathbf{E}_{\text{FDR}}$ with respect to the target plane.

$$\mathbf{E}_{\text{back}} = -S\{G(z - z_t) A \mathbf{E}_C(z_t)\} \quad (9)$$

where $S$ is symmetry operator about the target plane.

---

[1] We use relativistic units $(c = 1)$ and rational definitions of fields and charge, for instance $\nabla \mathbf{E} = \rho$, $e^2/(4\pi\hbar) = \alpha = 1/137$.

The FDR amplitude at $z = z_t$ is most conveniently calculated in the $(x, k_y)$ representation (7) :

$$\mathbf{E}_{\text{FDR}}(\omega, x, k_y, z_t) = -\frac{ie}{2}\theta(x-b) e^{-\mu|x|} e^{i\omega z_t/v} (i\,\text{sign}(x), \tau). \quad (10)$$

We have approximated $v$ by 1, except in the exponentials. Taking the Fourier transform in $k_x$, we obtain the full $\mathbf{k}_T$ representation which, for $b > 0$ (non-intercepting screen), reads

$$\mathbf{E}_{\text{FDR}}(\omega, \mathbf{k}_T, z_t) = -\frac{ie}{2(\mu + ik_x)} e^{-(\mu+ik_x)b} e^{i\omega z_t/v} (i, \tau). \quad (11)$$

Using (4) one obtains the well-known energy-angle spectrum of Diffraction Radiation (here $\mathbf{k}_T = \omega\vec{\theta}$):

$$\frac{dW}{\hbar d\omega d\Omega} = \frac{R^2}{\pi} |\mathbf{E}_{\text{FDR}}(\omega, \mathbf{R})|^2 = \frac{\alpha\omega^2}{4\pi^2\mu^2} \frac{q_0^2 + 2k_y^2}{q_0^2 + \mathbf{k}_T^2} e^{-2\mu b}. \quad (12)$$

*4.4. Generalization to two targets*

Let us now consider two successive targets $T_1$ (absorbing or reflecting) and $T_2$ (reflecting), located at $z_1$ and $z_2$ and of supporting functions $A_1$ and $A_2$. The field between $T_1$ and $T_2$ is

$$\mathbf{E}_{\text{betw}}(z) = \mathbf{E}_{\text{FDR}}^{(1)}(z) + \mathbf{E}_C(z), \quad (13)$$

with $\mathbf{E}_{\text{FDR}}^{(1)}(z)$ given by (8) with $z_t = z_1$. In the backward direction from the mirror $T_2$ one observes the reflection of $\mathbf{E}_{\text{betw}}$ by $T_2$:

$$\mathbf{E}_{\text{back 2}} = -S_2 \{G(z-z_2) A_2 \mathbf{E}_{\text{betw}}(z_2)\} \quad (14)$$

where $S_2$ is symmetry operator about the $T_2$ plane. Leaving aside the symmetry about the $T_2$ plane, the energy-angle distribution of $\mathbf{E}_{\text{back 2}}$ is determined by the $\mathbf{k}_T$ representation of the field

$$\mathbf{F} = A_2 \mathbf{E}_{\text{betw}}(z_2) = A_2 G(z_2 - z_1) \mathbf{E}_{\text{FDR}}^{(1)}(z_1) - \mathbf{E}_{\text{FDR}}^{(2)}(z_2), \quad (15)$$

where (8) has been used for target $T_2$ in the last term.

*4.5. Analytical results*

The analytical calculation starting from (6) are most easily done in the $(x, k_y)$ representation. Here we only give the final results for the same-side and opposite-side cases. The details of the derivation can be found in [12]. We take $z_1 = 0$, $z_2 = L$.

*4.5.1. Targets on the same side.*

Their supporting functions are $A_1(x, y) = \theta(x - b_1)$ and $A_2(x, y) = \theta(x - b_2)$ with $b_1, b_2 > 0$. The amplitude in the $\mathbf{k}_T$ representation is

$$\mathbf{F}(\omega, \mathbf{k}_T) = \frac{ie}{2}(i, \tau)\frac{e^{-(\mu+ik_x)b_2}}{\mu+ik_x}\{e^{i\omega L/v} - e^{i\omega L}[$$

$$e^{(\mu+ik_x)(b_2-b_1)}\exp\left(-iL\frac{k_T^2}{2\omega}\right)\frac{1}{2}\text{Erfc}\left[(2iL/\omega)^{-1/2}(b_2 - b_1 - k_x L/\omega)\right]$$

$$+\exp\left(iL\frac{\mu^2-k_y^2}{2\omega}\right)\frac{1}{2}\text{Erfc}\left[(2iL/\omega)^{-1/2}(b_1-b_2+i\mu L/\omega)\right]\}. \qquad (16)$$

The function $\text{Erfc}(z) = 2\pi^{-1/2}\int_z^{+\infty} dt\, e^{-t^2} = 1 - erf(z)$ is the complementary error function for complex argument. One can check that $\mathbf{F}(\omega, \mathbf{k}_T)$ is zero for $L \simeq 0$ and $b_1 < b_2$ (full shadowing) and is equal to $\mathbf{E}_{FDR}(b_2) - \mathbf{E}_{FDR}(b_1)$ (partial shadowing) for $L \simeq 0$ and $b_1 > b_2$.

*4.5.2. Targets on opposite sides.*
We have still $A_1(x,y) = \theta(x-b_1)$ with $b_1 > 0$ but $A_2(x,y) = \theta(b_2 - x)$ with $b_2 < 0$. We obtain

$$\frac{2}{ie}\mathbf{F}_{opp}(\omega, \mathbf{k}_T) = \frac{(-i,\tau)}{\mu - ik_x}e^{(\mu - ik_x)b_2}\exp(i\omega L/v) - \frac{(i,\tau)}{\mu + ik_x}e^{i\omega L}\{$$

$$e^{-(\mu + ik_x)b_1}\exp\left(-iL\frac{k_T^2}{2\omega}\right)\frac{1}{2}\text{Erfc}\left[(2iL/\omega)^{-1/2}(b_2 - b_1 - k_x L/\omega)\right]$$

$$+e^{-(\mu + ik_x)b_2}\exp\left(iL\frac{\mu^2 - k_y^2}{2\omega}\right)\frac{1}{2}\text{Erfc}\left[(2iL/\omega)^{-1/2}(b_1 - b_2 + i\mu L/\omega)\right]\}. \qquad (17)$$

*4.6. Numerical results*
Numerical calculations of the quantity $W = |F/e|^2$, which is proportional to the spectral-angular distribution and given by (16) or (17), were performed for the electron beam parameters indicated above for impact-parameter $b_1 = b_2 = 10$ mm and wavelength $\lambda = 12$ mm. First we consider the calculations for the case where targets are on the same side of the electron beam (see figure 6). In figure 15 the radiation angular distribution calculated with (16) for different values of the distance $L$ between absorber and conductive target for value interval of $L = 0 \sim 220$ mm is shown.
We can see the manifestation of the shadowing effect and the effect of the recovering of the electron field. Figure 16 shows the numerical results, calculated with (17), for the case where the targets are on opposite sides of the beam (figure 7). The intensity is much higher than in the same-side case and practically independent of $L$, which indicates that in this case the influence of shadowing is negligible. The theoretical plots of figures 15-16 are quite similar to the experimental ones of figures 13-14, in spite of the fact that concern only one wavelength. $\theta$ is the angle with respect to the direction of specular reflection on $T_2$. It corresponds to 85-$\theta$ (in degree) of figure 13 or 100-$\theta$ of figure 14.

**5. Discussion**
   The present experiment has demonstrated the feasibility of a direct observation of the shadowing of an electron electromagnetic field in a macroscopic mode. The used experimental method allowed us to make a basic analysis of this phenomenon. Unfortunately the experimental conditions in this experiment could not allow us to measure the *L*-dependence of the Coulomb field recovering up to $\gamma^2\lambda$ (*L* is the distance from the absorber). Nevertheless, in the observed *L* range, the expected properties of shadowing are clearly seen and the dependence of the radiation intensity on *L* and on the observation angle is in a good agreement with the theoretical one. A complementary confirmation of the pseudo-photons viewpoint is the identity of the shadow effects with absorbing and conductive screens.

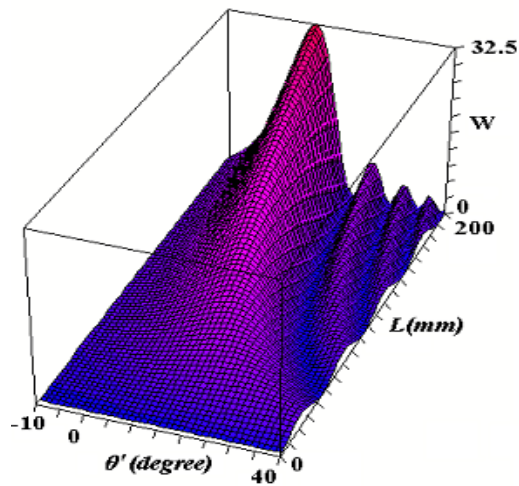 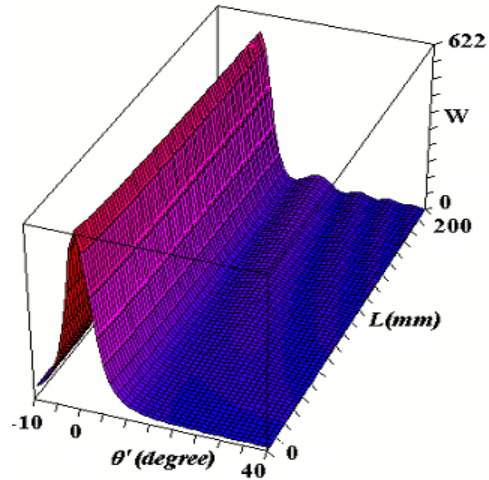

Figure 15. Calculated dependence of the radiation intensity on the observation angle θ' and distance $L$ between the targets for targets on the same side.

Figure 16. Same quantity as in Fig. 15, but for targets on opposite sides

We have presented the shadow effect from two points of view:

A) For the phenomenology, the electron field properties at $\gamma^2 \gg 1$ are close to those of an electromagnetic field in free space, therefore pseudo-photons are reflected by the mirror or absorbed in the absorber almost like real photons.

B) For the calculation, we have considered the shadowed electron field as a sum of the electron field in free space and the forward diffraction radiation (FDR) emitted from the target (or, if the electron crosses the target, forward transition radiation FTR).

The qualification "emitted from the target" applied to FDR or FTR may be confusing. In a traditional interpretation (for example in [6]), DR and TR from perfectly conducting targets are indeed considered as radiations emitted by surface currents. For non-perfectly conducting targets, surface currents are replaced by volume polarization currents. These surface or polarization currents are induced by the electromagnetic field of the relativistic electron. If we now take the viewpoint A and apply it to the absorber case, the electron electromagnetic field is evanescing in the target and no surface or polarization current may occur on the downstream side. One could then erroneously conclude (in this traditional interpretation) that no FDR or FTR is generated from an absorber.

The solution to this paradox is: either one uses Maxwell equations in matter (MEM), which include absorption, but do not take the polarizations currents as field sources, or one uses Maxwell equations in vacuum (MEV), which take the polarizations currents as sources but do not include absorption. It is therefore not correct to "screen" the field of the polarization currents. Let us show how these two formalisms are dual and give the same prediction in the case of forward optical transition radiation.

The Coulomb field can be decomposed as $F_C=F_u+F_d$, where $F_u$ is the retarded field of the upstream electron trajectory (suddenly stopped electron) and $F_d$ the retarded field of the downstream trajectory (suddenly accelerated electron). In the MEM formalism, FTR is generated by the current of the traveling electron after traversal of the screen [13]}. This current also re-creates the Coulomb field. The total field in the forward region is then $F_d = F_{FTR} + F_C$. Far enough from the trajectory, $F_{FTR} \approx F_d$. We can then say that FTR is generated during the transition of electron from an unstable "naked" state to the stable "dressed" state (here the term "transition radiation" seems particularly appropriate). A similar interpretation of this phenomenon can be found in [2]. Similarly, forward diffraction radiation can be viewed as a consequence of the transit from a "half-naked" state to the dressed state.

In the MEV formalism, $F_{FTR}$ is equal in the forward region to the field $F_m$ generated by the currents circulating on or under the upstream surface of the target. These currents are induced by $F_C$, which is nearly equal to $F_u$ in the target, and screen it, therefore $F_m \sim - F_u$ and the total forward field is $F_C + F_m = F_d$, as in the MEM formalism. Nevertheless the MEM approach is more intuitive for FTR.

Another example where the shadow effect is at work is Smith-Purcell (SP) radiation. One can consider the SP effect as a reflection of the pseudo-photons on the ridges of the grating. Then one expects that each ridge makes a shadow on the next one. The grating period, $\Lambda = \lambda/(1-v\cos\theta)$ is indeed smaller than $l_f = \lambda\gamma^2$ for a typical emission angle $\theta$. The shadow effect suggests a limit of the form $W \leq C \cdot \hbar L/(137 \cdot b^2)$ on the energy of Smith-Purcell radiation per electron [14,12]. $C$ is a numerical constant, $L$ the length of the electron path above the grating and $b$ the impact parameter, *i.e.*, the distance between the electron trajectory and the tops of the ridges. In fact, such a bound with the precise value $C=1/(2\pi)$ can be derived from unitarity and analyticity of the reflection coefficients and detailed bounds are obtained for fixed $\omega$ and fixed component $k_y$ of the photon momentum [15].

**Acknowledgment**

This work was partly supported by the warrant-order 1.226.08 of the Ministry of Education and Science of the Russian Federation and by the Federal agency for science and innovation, contract 02.740.11.0245